\documentclass[aps,amsmath,amssymb,nofootinbib]{revtex4}

\usepackage{graphicx}
\usepackage{graphics}
\usepackage{amsmath}
\usepackage{subfigure}

\begin{document}
\title{Phase Transitions Between Solitons and Black Holes in Asymptotically AdS/$\mathbb{Z}_k$ Spaces}

\author{Sean Stotyn}
\email{smastoty@sciborg.uwaterloo.ca}
\author{Robert Mann}
\email{rbmann@sciborg.uwaterloo.ca}

\affiliation{Department of Physics and Astronomy, University of Waterloo,\\
                   Waterloo, Ontario, Canada, N2L 3G1}

\begin{abstract}

We employ a thermodynamic analysis to determine the phase structure of Eguchi-Hanson solitons, Schwarzschild-AdS/$\mathbb{Z}_k$ black holes and thermal AdS/$\mathbb{Z}_k$.  The Euclidean actions are calculated by two equable means: the first uses the Eguchi-Hanson soliton as the thermal background while the second makes use of minimal boundary counterterms in the action necessary to render individual actions finite.  The Euclidean actions are then utilised to determine the phase structure in arbitrary odd dimension; it is found that there is a Hawking-Page phase transition and also a phase transition between the black hole and soliton.  There is found to be no smooth phase transition governed by an order parameter between AdS/$\mathbb{Z}_k$ and the soliton but nevertheless AdS/$\mathbb{Z}_k$ changes phase by tunneling to the lower energy soliton configuration.

\end{abstract}

\maketitle

\section{Introduction}

Arguably one of the deepest insights to emerge from string theory considerations in recent years is the AdS/CFT correspondence.  Holographically relating gravitational phenomena in the bulk to a conformal field theory living on the boundary, the correspondence has wide reaching application in furthering our understanding of either gravitational or particle physics that is otherwise difficult to calculate directly.  For instance, if it is possible to make calculations on the gravity side then one can infer information about the CFT and conversely if the CFT is calculable then one can infer information about the intractable gravity.  Of course there are also examples where both the gravity and field theory sides can be independently studied.

An example of this is the Hawking-Page phase transition between thermal AdS and the Schwarzschild AdS black hole \cite{Hawking:1982dh} and the corresponding confinement/deconfinement phase transition in the large $N$ limit of ${\cal N}=4$ super Yang-Mills theory \cite{Witten:1998qj}.  Another such example is the conjecture by Horowitz and Silverstein that a soliton configuration, a so-called `bubble of nothing' , is dual to closed string tachyon condensation \cite{Horowitz:2005vp,Horowitz:2006mr}.  Examples of such configurations are the AdS solitons found by Horowitz and Myers \cite{Horowitz:1998ha} and the Eguchi-Hanson solitons found by Clarkson and Mann \cite{Clarkson:2006zk} in 5 dimensions.  Both of these soliton configurations were shown to be the lowest energy solutions to the Einstein equations in their asymptotic classes and hence represent ground states.  The 5-dimensional Eguchi-Hanson soliton was subsequently obtained by Copsey \cite{Copsey:2007} using different methods, and higher odd-dimensional versions were obtained by Clarkson and Mann \cite{Clarkson:2005qx}.

Inspired by the AdS soliton being of lowest energy it was shown by Surya, Schleich and Witt in \cite{Surya:2001vj} that despite a lack of a phase transition between thermal AdS and AdS black holes with Ricci flat horizons \cite{Mann:1998tb,Birmingham:1998nr,Vanzo:1997gw} there is indeed a (somewhat different) phase transition between such black holes and the AdS soliton.  Furthermore, this transition was again shown to be dual to a confinement/deconfinement phase transition in the gauge theory.  Recently Hikida and Iizuka \cite{Hikida:2007pr,Hikida:2006qb} performed a detailed study in $D=3,5$ of the phase behaviour of the field theory dual to Eguchi-Hanson solitons (EHS), orbifolded AdS (OAdS) and orbifolded Schwarzschild-AdS black holes (OSAdS).  They have identified both the EHS and OAdS as corresponding to local vacua of the gauge theory with the EHS being the expected ground state and dual to closed string tachyon condensation.

This analysis did very little insofar as  the gravitational side of the correspondence is concerned. It is the aim of this paper to provide such an analysis in arbitrary odd dimension with a primary focus on $D=5,7,9$, as these dimensions are of specific relevance in superstring theory.

\section{Solitons and Black Holes} 

We start with the three known static, uncharged, asymptotically AdS/$\mathbb{Z}_k$ spacetimes in $D=2n+1$ dimensions whose metrics take the general form 
\begin{equation}
ds^2=-f_i(r)dt^2+\frac{dr^2}{f_i(r)h_i(r)}+\frac{r^2}{2n}\sum_{j=1}^{n-1}(d\theta_j^2+\sin^2\theta_jd\phi_j^2)+\left(\frac{r}{n}\right)^2h_i(r)\left(d\psi_i+\sum_{j=1}^{n-1}\cos\theta_jd\phi_j\right)^2
\end{equation}
where the subscript $i=s,a,b$ denotes the EHS, OAdS and OSAdS respectively and the angular coordinates take value in the range $\theta_j\in[0,\pi]$, $\phi_j\in[0,2\pi]$ and $\psi_i$ has period $\eta_i$.  For the EHS to be free of conical singularities we must impose $\eta_s=\frac{4\pi}k$ for some integer $k\ge3$ \cite{Clarkson:2005qx}.  The metric functions $f_i(r)$ and $h_i(r)$ take the following form for the three metrics:
\begin{eqnarray}
&&f_s(r)=1+\frac{r^2}{l^2} \>\>\>\>\>\>\>\>\>\>\>\>\>\>\>\>\>\>\>\>\>\>\>\>\>\>\>\>\>\>\>\>\>\>\>\>\>\>\>\> h_s(r)=1-\left(\frac{a}{r}\right)^{2n}\\
&&f_a(r)=1+\frac{r^2}{l^2} \>\>\>\>\>\>\>\>\>\>\>\>\>\>\>\>\>\>\>\>\>\>\>\>\>\>\>\>\>\>\>\>\>\>\>\>\>\>\>\>h_a(r)=1\\
&&f_b(r)=1-\frac{\mu}{r^{2n-2}}+\frac{r^2}{l^2} \>\>\>\>\>\>\>\>\>\>\>\>\>\>\>\>\>\>\>\>\>\>\>h_b(r)=1
\end{eqnarray}
where $l$ is related to the cosmological constant via $\Lambda=-\frac{n(2n-1)}{l^2}$, the soliton radius $a$ is restricted by the relation $a^2=l^2\left(\frac{k^2}{4}-1\right)$ as is explained in \cite{Clarkson:2006zk,Clarkson:2005qx} and $\mu$ is the mass parameter for the black hole.  For most of what follows, using $\mu$ is inconvenient so we introduce instead the horizon radius $r_+$ in terms of which $\mu=r_+^{2n-2}\left(1+\frac{r_+^2}{l^2}\right)$.  

Smooth instantons can be constructed for all three solutions by performing the Wick rotation $t\rightarrow i\tau$ and identifying $\tau$ with period $\beta_i$.  For the EHS and OAdS $\beta$ can take any value whereas for OSAdS the elimination of conical singularities requires
\begin{equation}
\beta_b=\frac{2\pi r_+}{n-1+n\frac{r_+^2}{l^2}}
\end{equation}

\section{Phase Transitions and Stability}

We are now ready to evaluate the Euclidean actions for the three solutions for use in thermodynamic analysis.  If one attempts to compute the action for a non-compact manifold one obtains an infinite answer due to the infinite volume, which is clearly undesirable.  We will employ two separate methods for rendering the action finite: the first uses a reference background as a subtraction and the second uses an insertion of boundary terms in the action that leave the equations of motion unchanged.  We will perform the former for arbitrary dimension and the latter for the specific case of $D=7$ as a consistency check.

The background subtraction method uses the Einstein-Hilbert action with the Gibbons-Hawking counterterm:
\begin{equation}
{\cal I}=-\frac1{16\pi G}\int_{\cal M}{d^Dx\sqrt{g}({\cal R}-2\Lambda)}-\frac1{8\pi G}\int_{\partial{\cal M}}{d^{D-1}x\sqrt{\gamma}(K-K_0)} \label{eq:action1}
\end{equation}
where $G$ is the D-dimensional Newton constant, $g$ is the determinant of the (Euclidean) metric on $\cal M$, $\cal R$ is the Ricci scalar, $\gamma$ is the determinant of the (Euclidean) metric on $\partial {\cal M}$ and $K$ and $K_0$ denote the trace of the extrinsic curvature of $\partial {\cal M}$ in the metric $g$ and the background metric respectively.  For the solutions we are interested in, the boundary terms vanish and the Ricci scalar is given simply by $R=-\frac{2n(2n+1)}{l^2}$ so the action reduces to
\begin{equation}
{\cal I}=\frac{n}{4\pi l^2 G}\int_{\cal M}{d^Dx\sqrt{g}}
\end{equation}
which is just a constant multiplying the infinite volume of the manifold $\cal M$.  To render the action finite, we need to appropriately subtract the background action by integrating $r$ out to some large value $R$, matching the periodicities of the angular coordinates and taking the limit as $R\rightarrow\infty$.   In \cite{Clarkson:2005qx} it was shown that in $D=5$ the EHS is perturbatively the solution of lowest energy.  Although no such analysis has been performed for the higher dimensional soliton solutions, we expect this result to remain valid and hence we use the EHS as the background.  Thus, we are ultimately interested in the relative action $I_i= {\cal I}_i-{\cal I}_s$.

To ensure the periodicities of the angular coordinates match up properly when taking the limit $R\rightarrow\infty$, the following matching conditions must be applied:
\begin{equation}
\>\>\>\>\>\>\>\>\>\>\>\>\>\>\>\>\beta_s=\beta_a \>\>\>\>\>\>\>\>\>\>\>\>\>\>\>\>\>\>\>\>\>\>\>\>\>\>\>\>\>\>\>\>\>\>\>\>\>\> \sqrt{h_s(R)}\eta_s=\eta_a
\end{equation}
\begin{equation}
\sqrt{f_s(R)}\beta_s=\sqrt{f_b(R)}\beta_b  \>\>\>\>\>\>\>\>\>\>\>\>\>\>\>\>\>\>\>\>\>\>\> \sqrt{h_s(R)}\eta_s=\eta_b
\end{equation}
With these conditions, the actions can be written in terms of $\beta_b$ and $\eta_s$, which are both known.  The final expressions for the background subtracted actions are
\begin{equation}
I_s=0
\end{equation}
\begin{equation}
I_a=\frac{2^{n-3}\pi^{n-1}}{l^2n^nkG}\beta_b a^{2n}\label{eq:Ia}
\end{equation}
\begin{equation}
I_b=\frac{2^{n-3}\pi^{n-1}}{l^2n^nkG}\beta_b\left[a^{2n}+r_+^{2n}\left(\frac{l^2}{r_+^2}-1\right)\right] \label{eq:Ib}
\end{equation}
We note that in choosing $n=1,2$ we are able to reproduce the results of \cite{Hikida:2006qb,Hikida:2007pr}, albeit in a slightly different way.

The other method we consider renders individual actions finite without invoking some non-unique background geometry.  It employs boundary terms in the action which leave the equations of motion the same but which have a non-trivial effect on the stress tensor.  The action we consider is
\begin{equation}
{\cal I}=-\frac1{16\pi G}\int_{\cal M}{d^Dx\sqrt{g}({\cal R}-2\Lambda)}-\frac1{8\pi G}\int_{\partial{\cal M}}{d^{D-1}x\sqrt{\gamma}K}+\frac1{8\pi Gl}\int_{\partial{\cal M}}{d^{D-1}x\sqrt{\gamma}\Theta} \label{eq:action2}
\end{equation}
which is very similar to (\ref{eq:action1}) but with two key distinctions.  First, there is no longer a background to subtract so $K_0$ is no longer present and the boundary contribution from the extrinsic curvature no longer vanishes.  Secondly, there is a new boundary term in which $\Theta$ is a dimensionless function of the boundary geometry; specifically $\Theta$ can only be a function of the boundary Riemann tensor, its contractions and their derivatives \cite{Henningson:1998gx,Balasubramanian:1999re,Mann:1999ms,Myers:1999}

An algorithm for calculating $\Theta$ for general $D$ in terms of a series expansion in the AdS radius was put forth in Ref. \cite{Kraus:1999di} and the explicit form in $D=3,5$ has been constructed in Ref. \cite{Balasubramanian:1999re}.  Furthermore, in Ref. \cite{Henningson:1998gx} it was shown that such a $\Theta$ indeed exists in $D=7$.  The series expansion in the AdS radius effectively amounts to a series expansion in inverse powers of the boundary radius so in what follows we will choose the convenient basis
\begin{equation}
\Theta=A+B\frac{l^2}{R^2}+C\frac{l^4}{R^4}
\end{equation}
where $R$ is the radius of the boundary, not the boundary Ricci scalar. A quick computation shows that in $D=7$ the extrinsic curvature for all three solutions is given by $K=\frac 6 l+\frac{2l}{R^2}-\frac{l^3}{4R^4}+{\cal O}(\frac 1 {R^6})$.  Bringing everything together, we find the actions (\ref{eq:action2}) for all three solutions of interest are finite provided $A=5$, $B=\frac 5 2$ and $C=-\frac 5 8$ which agrees exactly with what one gets from the algorithm of Ref. \cite{Kraus:1999di}.  This ``quick and dirty" algorithm to find $\Theta$ by writing it as a power series in inverse powers of the boundary radius and then finding the coefficients that render the actions finite generalizes to any number of dimensions.  A more thorough discussion of how to write down $\Theta$ in any dimension was given in Ref. \cite{Lorenzo:2003}.

The resulting actions are given explicitly by
\begin{equation}
{\cal I}_s=-\frac{\pi^2\beta_b}{216kl^2G}(8a^6+5l^6)
\end{equation}
\begin{equation}
{\cal I}_a=-\frac{5\pi^2l^4\beta_b}{216kG}
\end{equation}
\begin{equation}
{\cal I}_b=\frac{\pi^2\beta_b}{216kl^2G}(l^2r_+^4-r_+^6-5l^6)
\end{equation}
It is easily verified that defining $I_a\equiv {\cal I}_a-{\cal I}_s$ and $I_b\equiv {\cal I}_b-{\cal I}_s$ reproduces (\ref{eq:Ia}) and (\ref{eq:Ib}) respectively for the case $n=3$ so both methods are seen to produce the same physics as they should; for related analyses carried out in $D=3,5$ see Refs. \cite{Hikida:2007pr,Balasubramanian:1999re}.  For the sake of generality in what follows we use the background subtracted actions (\ref{eq:Ia}) and (\ref{eq:Ib}), the only difference being the zero of the action which we choose to be the EHS.  We also emphasize that there are two independent thermodynamic parameters, namely $r_+$ and $l$, and it should be understood that $\beta_b=\beta_b(r_+,l)$ and $a=a(l)$; $n$ and $k$ are held fixed as they define the dimension and topology respectively and cannot change\footnote{In \cite{Hikida:2007pr} it was shown that in 3 dimensions ($n=1$) there is a topological transition from spaces with orbifold parameter $k'$ to spaces with 
orbifold parameter $k<k'$ eventually ending in the stable vacuum of $AdS_3$; this topological transition is very special to $D=3$ and is not present in the higher dimensions we consider in this paper.}.  

We make note of the fact that the OAdS action (\ref{eq:Ia}), and hence the free energy ${\cal F}_a=\beta_b^{-1} I_a$, is always positive and so the OAdS geometry is always an unstable phase.  That is to say there is no order parameter that can be tuned to bring about a continuous phase transition between the OAdS and EHS geometries.  This supports the notion mentioned earlier that both the EHS and OAdS solutions are vacuum states with the EHS being the lowest energy ground state.  In fact, the thermodynamic energy, $\langle E\rangle=\frac{\partial I}{\partial \beta_b}$, of OAdS is given by
\begin{equation}
\langle E\rangle_a=\frac{2^{n-3}\pi^{n-1}a^{2n}}{l^2n^nkG}
\end{equation}
The OAdS geometry then lowers its free energy to zero by tunneling to the EHS geometry at the rate $\Gamma\sim e^{-\beta_b \langle E\rangle_a}=e^{-I_a}$.  The entropy, $S=\beta_b\langle E\rangle-I$, and the heat capacity, $C=-\beta^2\frac{\partial^2 I}{\partial \beta^2}$, both vanish for OAdS as they ought to for a horizonless spacetime.

To analyse the thermodynamic behaviour of the black hole it is convenient to introduce the Hawking-Page action
\begin{equation}
I_{HP}\equiv {\cal I}_b-{\cal I}_a=I_b-I_a \label{eq:HP}
\end{equation}
so that we can write $I_b=I_a+I_{HP}$.  The thermodynamic energy of OSAdS is found to be given by
\begin{equation}
\langle E\rangle_b=\langle E\rangle_a+\frac{2^{n-3}\pi^{n-1}}{n^nkG}(2n-1)\mu \label{eq:Eb}
\end{equation}
where $\mu=\mu(r_+,l)$ is the black hole mass parameter defined previously and we immediately see the utility of (\ref{eq:HP}): the first term in (\ref{eq:Eb}) is due to $I_a$ and the second term is due to $I_{HP}$.  We have already shown that the entropy and heat capacity vanish for $I_a$ so we conclude that these quantities are governed by the Hawking-Page action alone.  As expected for the entropy, we find
\begin{equation}
S_b=\frac{2^{n-1}\pi^n}{n^nkG}r_+^{2n-1}=\frac{A_b}{4G}
\end{equation}
where $A_b$ is the area of the black hole horizon.  Similarly, the heat capacity is found to be
\begin{equation}
C=\frac{(2n-1)2^n\pi^{n+1}r_+^{2n}}{\beta_bn^nkG\left(n\frac{r_+^2}{l^2}-(n-1)\right)}
\end{equation}
which diverges at $r_+=r_0$ where $r_0^2=\frac{n-1}n l^2$ signifying a first order phase transition.  This is precisely where $\frac{\partial\beta_b}{\partial r_+}=0$ and hence signifies the transition between small and large black holes with $\beta_b(r_0,l)$ giving the lowest possible temperature an OSAdS black hole can have.

Next we want to look at the thermodynamic stability of the black hole.  We have already concluded that OAdS is unstable and tunnels to EHS since its free energy is always positive.  We thus would like to know if the black hole action (\ref{eq:Ib}) can be negative, that is to say if OSAdS can be thermodynamically favoured over EHS.  To establish this, we need to find some $r_+=r_1$ such that $I_b(r_1,l)=0$ signifying the cross over from positive to negative free energy.  If we define $\chi_1\equiv \frac{r_1^2}{l^2}$ then we need to solve
\begin{equation}
\left(\frac{k^2}4-1\right)^n-\chi_1^{n-1}(\chi_1-1)=0. \label{eq:fchi}
\end{equation}
This equation is a simple polynomial of degree $n$ and it turns out we are guaranteed at least one real solution due to the mean value theorem: the action evaluated at $r_+=r_0$ is positive since
\begin{equation}
\left(\frac{k^2}4-1\right)^n-\left(\frac{r_0}l\right)^{2n-2}\left(\frac{r_0^2}{l^2}-1\right)=\left(\frac{k^2}4-1\right)^n+\frac{(n-1)^{n-1}}{n^n}>0
\end{equation}
whereas the action for sufficiently large $r_+=R$ is negative since 
\begin{equation}
\left(\frac{k^2}4-1\right)^n-\left(\frac{R}l\right)^{2n-2}\left(\frac{R^2}{l^2}-1\right)\rightarrow-\left(\frac{R}l\right)^{2n}<0.
\end{equation}
We therefore know that such an $r_1$ exists and furthermore that $r_1>r_0$.    In Fig. 1 we have plotted (\ref{eq:fchi}) for various values of $n$ and $k$; it is where the coloured lines intersect the x-axis that gives the position of $\frac{r_1^2}{l^2}$.  In plot (a), $n=3$ is held fixed while $k$ is allowed to vary: it is clearly seen that as one increases $k$, the value of $r_1$ increases as well.  In plot (b), $k=3$ is held fixed while $n$ varies: in contrast to increasing $k$, as $n$ increases $r_1$ decreases.  In both of these plots $\frac{r_0}{l}<1$ so our assertions that $r_1$ exist and that $r_1>r_0$ are seen to hold true.  Furthermore, we can conclude that since $r_1$ decreases with increasing $n$ there is `less room' for phase transitions to occur in higher dimensions whereas since $r_1$ increases with increasing $k$ there is `more room' for phase transitions to occur in more highly orbifolded topologies.

\begin{figure}
    \centering
     \subfigure[]{\includegraphics[width=3.5 in]{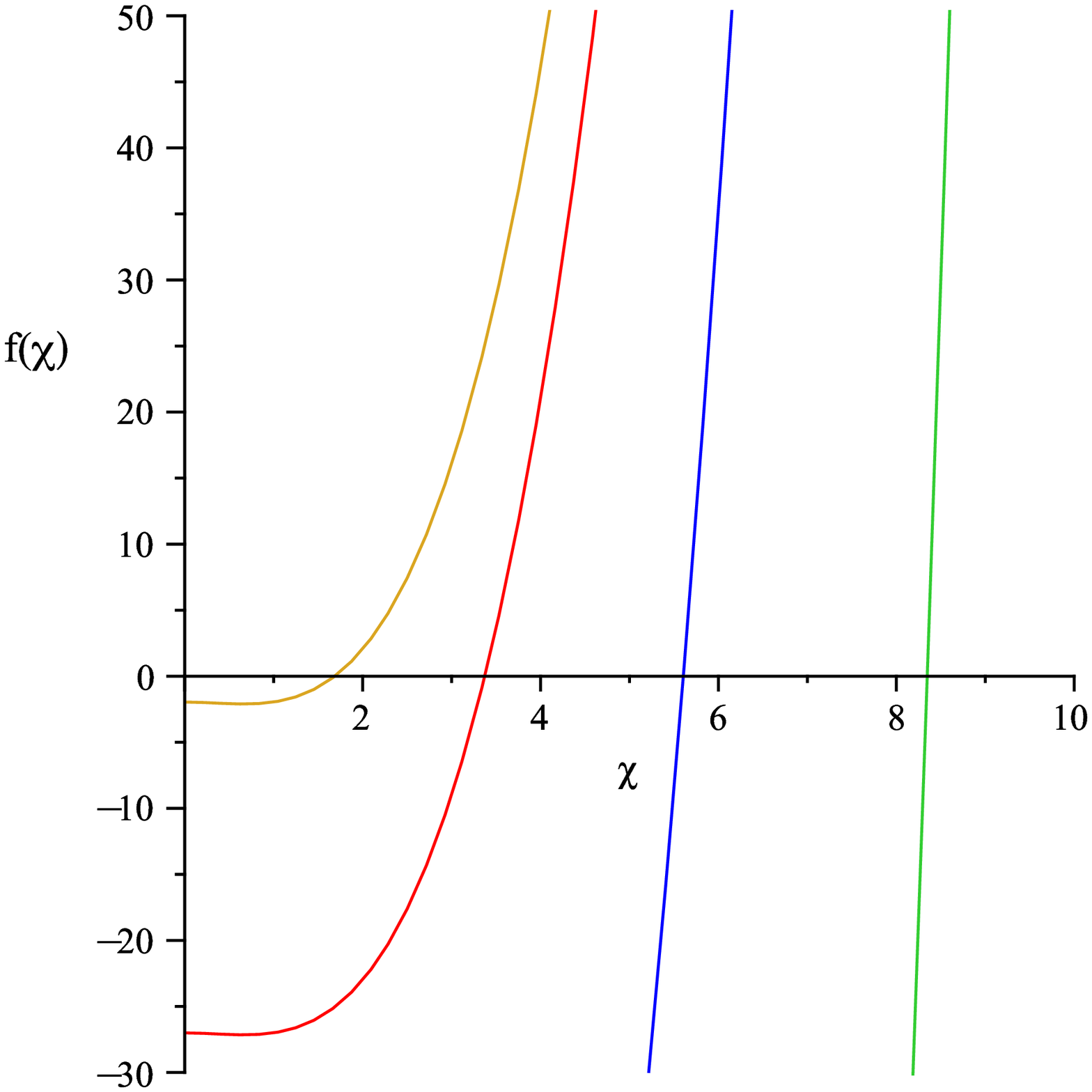}}
     \subfigure[]{\includegraphics[width=3.5 in]{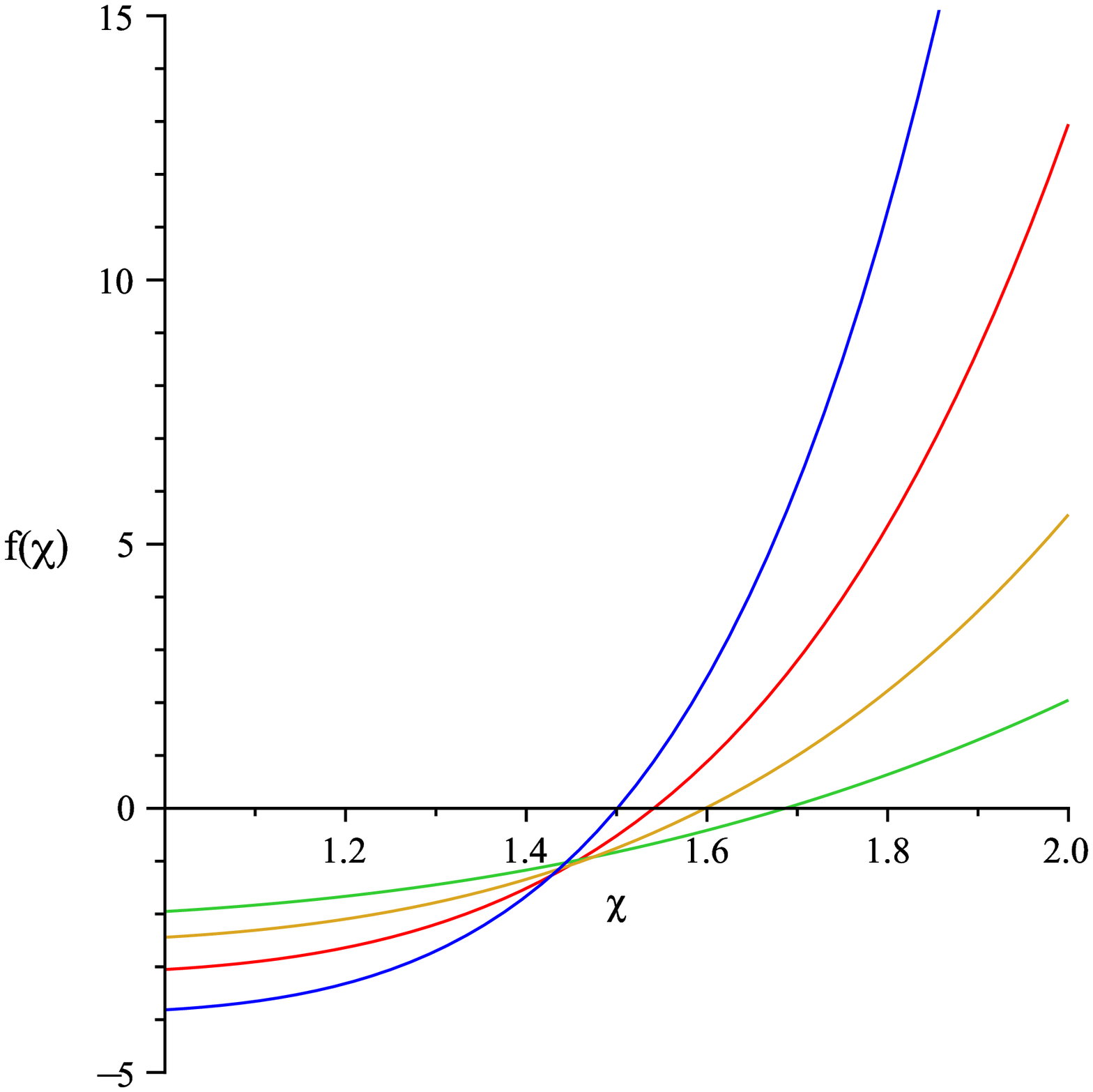}}
     \caption{(a) A plot of $f(\chi)$ where $n=3$ is held fixed and $k$ is varied: orange is $k=3$, red is $k=4$, blue is $k=5$ and green is $k=6$.  (b) A plot of $f(\chi)$ where $k=3$ is held fixed and $n$ is varied: blue is $n=6$, red is $n=5$, orange is $n=4$ and green is $n=3$.}
     \label{fig:plots}
\end{figure}

Since the free energy of OAdS is positive while that of OSAdS can be positive or negative, there exists another critical point, $r_+=r_{HP}$, above which the free energy of OSAdS is higher than OAdS and below which it is lower:  this happens at $r_{HP}=l$ and this is, not surprisingly, where the Hawking-Page action (\ref{eq:HP}) vanishes.  It is easily verified that $r_0<r_{HP}<r_1$.

Large black holes in the range $r_+>r_0$ have positive specific heat but this does not necessarily mean that they will be in a stable equilibrium with thermal radiation.  Indeed we can see from the arguments above that in the range $r_0<r_+<r_1$ the black hole has a higher free energy than the soliton and will thus lower its free energy by evaporating into thermal OAdS and subsequently tunneling to the soliton.  Furthermore, in the temperature range $\beta_b(r_{HP}<r_+<r_1)$ the large black hole has lower free energy than thermal OAdS but higher free energy than the EHS so it is more likely for thermal OAdS to tunnel to a large black hole configuration, which would then ultimately tunnel to the soliton, than it would be to tunnel to the soliton directly.  In the temperature range $\beta_b(r_+>r_1)$ the large black hole has the lowest free energy and hence thermal OAdS and the EHS will tend to tunnel to the black hole configuration.  Such black holes have  
 positive specific heat and are thermodynamically favoured and hence are genuinely in thermal equilibrium.  Small black holes on the other hand, corresponding to the range $r_+<r_0$, possess a negative specific heat and hence they are thermodynamically unstable and will radiate to thermal OAdS and eventually tunnel to the soliton.

\section{Concluding Remarks}

In this paper we have studied the thermodynamic behaviour of Eguchi-Hanson solitons and orbifolded AdS and SAdS spaces, generalising the work in Refs. \cite{Hikida:2007pr,Hikida:2006qb} to arbitrary odd dimension.  We have given a more thorough analysis on the gravity side than has previously been given for these spaces and we have left discussions related to the holographic dual description to these references and the appropriate references therein as the details are rigorously worked out.  The gravity description presented here very strongly supports the idea that OAdS and EHS are both (at least local) vacua with the EHS being the ground state.  Indeed OAdS is the end point of black hole evaporation, signifying a local vacuum, but OAdS has a positive free energy and hence tunnels to the zero free energy EHS configuration. This indicates that the EHS is at least a lower energy local vacuum than OAdS.  This is consistent with previous results \cite{Clarkson:2006zk} that demonstrated that in $D=5$ EHS spacetimes  are perturbatively of lowest energy 
in their topological class. While such an analysis remains to be done for the EHS in higher dimensions we anticipate the same outcome based on the results we have obtained.

We have shown that the expected results for the entropy and the heat capacity are obtained for the OSAdS solution and have uncovered a richer thermodynamic behaviour than in pure AdS spaces.  Namely all of the general elements of the usual Hawking-Page analysis are present, such as the existence of large and small black holes and a range of large black holes that are not stable despite the positivity of their specific heat but the effect of the soliton as the thermal background is to make wider the range of unstable large black holes.  This is simply due to the fact that there is a wider range of large black holes for which the free energy is still positive after dropping below the free energy of OAdS.

\section*{Acknowledgements}

This work was supported by the Natural Sciences and Engineering Research Council of Canada and the government of Ontario

\end{document}